\documentclass[12pt]{article}
\usepackage{graphicx}
\usepackage{epsfig}
\newcommand{\beq}{\begin{equation}}
\newcommand{\eeq}{\end{equation}}   
\newcommand{\beqn}{\begin{eqnarray}}
\newcommand{\eeqn}{\end{eqnarray}}  
\begin{document} 
 
\title{Form factors of the $K_{\mu3}$ decay in the spectator approximation 
of the Standard Model}

\author{V.P. Efrosinin, A.N. Khotjantsev\\
Institute for Nuclear Research RAS\\
60 October Revolution Pr. 7A, Moscow 117312, Russia}

\date{}
\renewcommand {\baselinestretch} {1.3}

\maketitle
\begin{abstract}

 We have calculated the  parameter $\xi(0)=f_-(0)/f_+(0)$ of the 
 decay $K^+ \to \pi^0\mu^+\nu_{\mu}$ in the spectator approximation of the
 Standard Model. It was obtained that  $\xi(0) \simeq -0.02$.   
\end{abstract}
The semileptonic decays of kaons afford an interesting opportunity  to study 
the quark currents. The isotopic symmetry of the strong interactions is 
successfully used for calculations of such processes. This appears to be 
an illustration of  a definite factorization of the  strong and weak 
interactions at low energies. This circumstance allows us to reveal mechanisms 
of the quark currents outside the framework of phenomenological approach. 
The semileptonic decay $K^+ \to \pi^0\mu^+\nu_{\mu}$~($K_{\mu 3}$), well 
studied experimentally, offers an interesting opportunity for such study.
  
\begin{figure}[ht]
\begin{center}
\epsfig{file=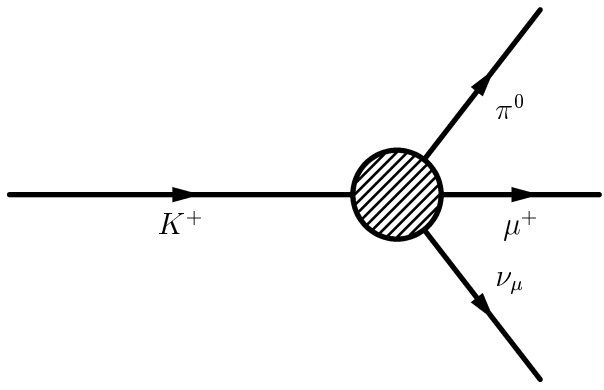,width=8cm}
\caption{The tree diagram of the $K_{\mu3}$ decay.}
\label{fig:fig1}
\end{center}
\end{figure}

We consider the  $K_{\mu 3}$ decay (Fig.~\ref{fig:fig1})
\begin{eqnarray}
\label{eq:Kpimunu}
K^+(p_K) \to \pi^0(p_{\pi})+\mu^+(p_{\mu})+\nu_{\mu}(p_{\nu}).
\end{eqnarray}
The tree diagram of this decay is shown in (Fig.~\ref{fig:fig1}).
The general structure of the Lagrangian of this decay is given by 
\begin{eqnarray}
\label{eq:L_W}
L_W=-\frac{G_F}{\sqrt{2}}sin\theta_C J^{+\mu}_{l} J_{h\mu},
\end{eqnarray}
where $J^{+\mu}_l$ is the weak leptonic current. It reads
\begin{eqnarray}
\label{eq:Jl}
J^{+\mu}_l=\bar u(p_{\nu})\gamma^{\mu}(1-\gamma_5)v(p_{\mu}),
\end{eqnarray}
and the phenomenological matrix element of the strangeness changing vector
current $J_{h\mu}$ can be expanded in the following form
\begin{eqnarray}
\label{eq:Jh}
J_{h\mu}&=&\langle\pi^0|\bar s\gamma_{\mu}u|K^+\rangle
\nonumber\\
&=&\frac{1}{\sqrt{2}}[f_+(t)(p_K+p_{\pi})_{\mu}+f_-(t)(p_K-p_{\pi})_{\mu}],
\end{eqnarray}
where $t=(p_K-p_{\pi})^2$ is  the four-momentum transfer  squared of mesons to 
the leptons.
  
It is accepted to consider that in this case only vector part of the weak 
hadronic current works. The matrix element of the axial-vector current is 
a pseudovector. It is impossible to construct a pseudovector from two momenta. 
In the case of the exact SU(3) symmetry,  $f_+(0)=1$ and $f_-(0)=0$. The 
physical situation corresponds to the disturbed SU(3) symmetry. The form 
factors $f_{\pm}(t)$ are represented as
\begin{eqnarray}
\label{eq:f+t}
f_{\pm}(t)=f_{\pm}(0)(1+\lambda_{\pm}\frac{t}{m^2_{\pi^+}}),
\end{eqnarray}
where parameter $\lambda_+\simeq 0.028$~\cite{Eidelman:2004wy}.

Parameter $\xi(0)$ is determined as the ratio of the two vector form factors 
\begin{eqnarray}
\label{eq:ksi}
\xi(0)=f_-(0)/f_+(0).
\end{eqnarray}
  
The value of  $\xi(0)$ can be obtained by three different methods:
by measuring the muon polarization, by studying the Dalitz plot of the 
$K_{\mu 3}$ decay, and also from  the $\Gamma(K_{\mu 3})/\Gamma(K_{e3})$ 
ratio~\cite{Eidelman:2004wy}. The consistent experimental values of 
$\xi\sim -0.1$ are obtained from $\lambda_+$, $\lambda_0$ and 
$\lambda_+$ - $\lambda_0$ measurement~\cite{Yushchenko:2003xz} and 
from the ratio $\Gamma(K_{\mu 3})/\Gamma(K_{e3})$~\cite{Horie:2001th}. 
The results of these experiments demonstrate a recovery of the  SU(3) symmetry 
because $\xi(0)\to 0$. But the mechanism of such recovery is not yet well
understood.

In the Standard Model~(SM), the amplitude of the $K_{\mu 3}$ decay  is given by
\begin{eqnarray}
\label{eq:T}
T&=&-\frac{G_F}{2}sin\theta_C \bar u(p_{\nu})\gamma^{\mu}(1-\gamma_5)
v(p_{\mu})[f_+(t)(p_K+p_{\pi})_{\mu}+f_-(t)(p_K-p_{\pi})_{\mu}]
\nonumber\\
&=&-G_Fsin\theta_C f_+\bar u(p_{\nu})[\gamma^{\mu}(1-\gamma_5)
(p_K)_{\mu}-\frac{1}{2}(\xi(t)-1)m_{\mu}(1+
\gamma_5)]v(p_{\mu}),
\end{eqnarray}
where $\theta_C$ is the Cabibbo angle, and $G_F$ is the Fermi coupling 
constant.

The amplitude of $K_{\mu 3}$ decay (\ref{eq:T})
in the $K^+$ rest frame can be represented as
\begin{eqnarray}
\label{eq:TT}
T=-G_Fsin\theta_C f_+\bar u(p_{\nu})[\gamma^0(1-\gamma_5)
m_K-\frac{1}{2}(\xi(t)-1)m_{\mu}(1+
\gamma_5)]v(p_{\mu}).
\end{eqnarray}

\begin{figure}[ht]
\begin{center}
\epsfig{file=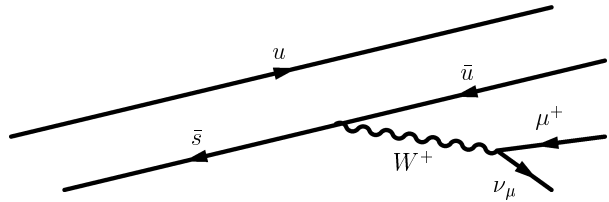,width=8cm}
\caption{The  diagram of the  $K_{\mu3}$ decay at quark level.}
\label{fig:fig2}
\end{center}
\end{figure}

In the SM the main contribution to amplitude (\ref{eq:T}) of the $K_{\mu 3}$ 
decay at the  quark level is described  by diagram (Fig.~\ref{fig:fig2}).
 In the spectator model, we consider below, the strange quark $\bar{s}$ 
interacts with the $W^+$ boson and transforms into  $\bar{u}$ quark. 
The $u$ quark continues independent motion. The subsequent strong 
interaction in a final state and fragmentation into a pion do not 
influence an angular distribution of leptons and the muon polarization. 
We do not include in this approach possible contributions to 
$K_{\mu3}$ decay from various extensions of the SM, because these non-SM 
models  are strongly constrained from the experimental  search for T-violation 
in this decay~\cite{Abe:2004px}.
 
The amplitude of the diagram (Fig.~\ref{fig:fig2}) in the spectator model can 
be written as
\begin{eqnarray}
\label{eq:T1}
T^1=-\frac{G_F}{\sqrt{2}}sin\theta_C\bar v(p_s)\gamma_{\mu}(1-\gamma_5)
v(p_u)\bar u(p_{\nu})\gamma^{\mu}(1-\gamma_5)v(p_{\mu}),
\end{eqnarray}
where $p_s$ and $p_u$ are the momenta of $\bar s$ and $\bar u$ quark, 
respectively.

These objects $\bar s$ and $\bar u$ can be current quarks, but we assume that 
they are constituent quarks. Such interpretation is proven due to the 
experimental test of a number of relations between amplitudes and cross 
sections. These ratios are derived in the framework of the impulse 
approximation of the additive quark model~\cite{Shekhter:1980zq}. In a simple 
approach~\cite{Efrosinin:1983zg}, there is a prediction for the 
existence of a soft scalar interaction of constituent quarks in a hadron, 
and the spin-spin interaction can be large. For example, it results in 
a considerable difference of masses of $\pi$ and $\rho$ mesons. 
We assume  that constituent quarks are non-relativistic. Taking into 
account the short-distance nature of the  weak interaction, we can
use an impulse approximation to interacting quarks within the framework 
of the spectator model. 

As it was mentioned above, only  vector part of the weak hadronic current 
contributes to the $K_{\mu3}$ decay. We will further use the Gordon identity 
for the vector current of the constituent $\bar s$ and $\bar u$ quarks
\begin{eqnarray}
\label{eq:gor}
\bar v(p_s)\gamma_{\mu}v(p_u)=\frac{1}{m_s+m_u}\bar v(p_s)[-(p_s+
p_u)_{\mu}+\sigma_{\mu \nu}(p_s-p_u)^{\nu}]v(p_u),
\end{eqnarray}
where
\begin{eqnarray*}
\sigma_{\mu \nu}=\frac{1}{2}(\gamma_{\mu}\gamma_{\nu}-\gamma_{\nu}
\gamma_{\mu}),
\end{eqnarray*}
$m_s$ is the rest mass of the constituent $\bar s$ quark, and $m_u$ 
is the rest mass of the constituent $\bar u$ quark in pion. 
It follows from Eq.~(\ref{eq:gor}) that  the tensor contribution should be 
added. However, the experimental study of the decay 
$K^{\pm}\to {\pi}^0 e^{\pm}\nu$ strongly constraints the contribution of 
the tensor  interaction to a few per cent level at 
(90\% C.L.)~\cite{Yushchenko:2004zs,Shimizu:2000im}. 
Therefore, we omit tensor coupling in our approach. 
So, using  Eq.~(\ref{eq:gor}), we can present amplitude (\ref{eq:T1}) as
\begin{eqnarray}
\label{eq:TT1}
T^1=\frac{G_F}{\sqrt{2}}sin\theta_C\frac{1}{m_s+m_u}\bar v(p_s)(p_s+
p_u)^{\mu}v(p_u)\bar u(p_{\nu})\gamma_{\mu}(1-\gamma_5)v(p_{\mu}).
\end{eqnarray}
In the spectator model
\begin{eqnarray}
\label{eq:psmu}
(p_s)^{\mu}=(p_u)^{\mu}+(p_{\mu})^{\mu}+(p_{\nu})^{\mu},
\end{eqnarray}
and
\begin{eqnarray}
\label{eq:TT11}
T^1&=&\frac{G_F}{\sqrt{2}}sin\theta_C\frac{1}{m_s+m_u}\bar v(p_s)(2 p_s-
p_{\mu}
-p_{\nu})^{\mu}v(p_u)\bar u(p_{\nu})\gamma_{\mu}(1-\gamma_5)v(p_{\mu})
\nonumber\\
&=&\frac{G_F}{\sqrt{2}}sin\theta_C\frac{\bar v(p_s)v(p_u)}{m_s+m_u}
\bar u(p_{\nu})(\gamma_{\mu}(1-\gamma_5)2 (p_s)_{\mu}
+m_{\mu}(1+\gamma_5))v(p_{\mu}).
\end{eqnarray}

In the rest frame of $K^+$ we make the assumption that $p_s=(m_s,0)$.
Then, from expression~(\ref{eq:TT11}) 
\begin{eqnarray}
\label{eq:TT12}
T^1=\frac{G_F}{\sqrt{2}}sin\theta_C\frac{\bar v(p_s)v(p_u)}{m_s+m_u}
\bar u(p_{\nu})(\gamma^0(1-\gamma_5)2 m_s
+m_{\mu}(1+\gamma_5))v(p_{\mu}).
\end{eqnarray}
The amplitude (\ref{eq:TT11}) must be multiplied by the factored contribution
of strong interaction of quarks in the final state. Then, it corresponds 
to the amplitude (\ref{eq:T}), and it is correct at small $t$. Thus, the ratio
of scalar and vector parts in  (\ref{eq:TT}) is equal to the same ratio in 
the expression (\ref{eq:TT12})
\begin{eqnarray}
\label{eq:TT13}
&&\frac{G_F sin\theta_C f_+\bar u(p_{\nu})(1/2)(\xi-1)
m_{\mu}(1+\gamma_5))v(p_{\mu})}
{-G_F sin\theta_C f_+\bar u(p_{\nu})\gamma^0(1-\gamma_5) m_K
v(p_{\mu})}\nonumber\\
&=&\frac{(G_F/\sqrt{2})sin\theta_C(\bar v(p_s)v(p_u)/(m_s+m_u))
\bar u(p_{\nu})m_{\mu}(1+\gamma_5))v(p_{\mu})}
{(G_F/\sqrt{2})sin\theta_C(\bar v(p_s)v(p_u)/(m_s+m_u))
\bar u(p_{\nu})\gamma^0(1-\gamma_5) 2 m_s v(p_{\mu})}.
\end{eqnarray}
From (\ref{eq:TT13}) we obtain
\begin{eqnarray}
\label{eq:m_s}
m_s=\frac{m_K}{1-\xi}.
\end{eqnarray}

Such spectator approximation will be correct only for a meson with
one heavy quark. For example, for the $\pi^+ \to \pi^0 e^+ \nu_e$ decay 
(exact $SU(2)$ symmetry, $\xi(0) = 0$) this approximation is not justified. 
Therefore, expression (\ref{eq:m_s}) has no sense for the precise unitary 
symmetry and can not be extrapolated to the $\pi^+ \to \pi^0 e^+ \nu_e$ 
decay. It is important that the constituent  mass of the strange quark is used 
in this formula.  
 
Let us consider a simple non-relativistic model of constituent quarks
with a spin-spin interaction for mesons \cite{Efrosinin:1983zg}:
\begin{eqnarray}
\label{eq:vs}
V_s=\beta_M\frac{{\bf S}_1 {\bf S}_2}{m_1 m_2},
\end{eqnarray}
where $V_s$ is the spin-spin interaction, ${\bf S}_i$ is the spin of 
constituent quark, $m_i$ is the mass of constituent quark, $\beta_M$ is 
the constant of the spin-spin interaction, 
\begin{eqnarray}
\label{eq:mpmr}
m_{\pi}&=&2m_u+E_0^M-(3/4)\beta_M m_u^{-2},\nonumber\\
m_{\rho}&=&2m_u+E_0^M+(1/4)\beta_M m_u^{-2},\nonumber\\
m_K&=&m_u+m_s+E_0^M-(3/4)\beta_M m_u^{-1}m_s^{-1},\nonumber\\
m_{K^*}&=&m_u+m_s+E_0^M+(1/4)\beta_M m_u^{-1}m_s^{-1}.
\end{eqnarray}
Here, $E_0^M$ is the scalar interaction, which is supposed to be small.
Then, from (\ref{eq:mpmr}) we have
\begin{eqnarray}
\label{eq:muq}
m_u\simeq\frac{m_{\pi}+3m_{\rho}}{8}\simeq~0.305~{\rm GeV},
\end{eqnarray}
\begin{eqnarray}
\label{eq:muq1}
m_u+m_s\simeq\frac{m_K+3m_{K^*}}{4}\simeq~0.792~{\rm GeV}.
\end{eqnarray}
Using equations (\ref{eq:m_s},\ref{eq:muq},\ref{eq:muq1}) one can obtain
\begin{eqnarray}
\label{eq:xi1}
\xi=1-\frac{m_K}{m_s}\simeq~-0.02.
\end{eqnarray}
The accuracy of $\xi$ in this approach is about 0.1 and 
determined by the uncertainty of the constituent s-quark mass of 
the model~\cite{Efrosinin:1983zg}. From the expression~(\ref{eq:m_s}), 
the mechanism of the recovery of the unitary symmetry relative to 
the value $\xi$ is seen. The extraction of $\xi(0)$ with small uncertainty 
is possible in the experiment E246~\cite{Abe:2003wi}, where  the normal 
polarization of the muon in the $K_{\mu 3}$ decay is expected to be 
measured with high statistics and small systematic errors. The further precise 
measurement of $\xi$ in the $K_{\mu 3}$ decay  will allow us to establish 
the mechanism of quark currents of week interactions.

\end{document}